\documentclass[twocolumn,aps,prb,showpacs,preprintnumbers,amsmath,amssymb, superscriptaddress]{revtex4-1}
\pdfoutput=1

\usepackage{bm}
\usepackage{graphicx}
\usepackage{color,xspace}

\begin{document}

\title{Disentangling the surface and bulk electronic structures of LaOFeAs}

\author{P. Zhang}\thanks{These authors contributed equally to this work.}
\affiliation{Beijing National Laboratory for Condensed Matter Physics, and Institute of Physics, Chinese Academy of Sciences, Beijing 100190, China}
\author{J. Ma}\thanks{These authors contributed equally to this work.}
\affiliation{Beijing National Laboratory for Condensed Matter Physics, and Institute of Physics, Chinese Academy of Sciences, Beijing 100190, China}
\author{T. Qian}
\affiliation{Beijing National Laboratory for Condensed Matter Physics, and Institute of Physics, Chinese Academy of Sciences, Beijing 100190, China}
\affiliation{Collaborative Innovation Center of Quantum Matter, Beijing, China}
\author{Y. G. Shi}
\affiliation{Beijing National Laboratory for Condensed Matter Physics, and Institute of Physics, Chinese Academy of Sciences, Beijing 100190, China}
\author{A. V. Fedorov}
\affiliation{Advanced Light Source, Lawrence Berkeley National Laboratory, Berkeley, California 94720, USA}
\author{J. D. Denlinger}
\affiliation{Advanced Light Source, Lawrence Berkeley National Laboratory, Berkeley, California 94720, USA}
\author{X. X. Wu}
\affiliation{Beijing National Laboratory for Condensed Matter Physics, and Institute of Physics, Chinese Academy of Sciences, Beijing 100190, China}
\author{J. P. Hu}
\affiliation{Beijing National Laboratory for Condensed Matter Physics, and Institute of Physics, Chinese Academy of Sciences, Beijing 100190, China}
\affiliation{Collaborative Innovation Center of Quantum Matter, Beijing, China}
\affiliation{Department of Physics, Purdue University, West Lafayette, Indiana 47907, USA}
\author{P. Richard}\email{p.richard@iphy.ac.cn}
\affiliation{Beijing National Laboratory for Condensed Matter Physics, and Institute of Physics, Chinese Academy of Sciences, Beijing 100190, China}
\affiliation{Collaborative Innovation Center of Quantum Matter, Beijing, China}
\author{H. Ding}\email{dingh@iphy.ac.cn}
\affiliation{Beijing National Laboratory for Condensed Matter Physics, and Institute of Physics, Chinese Academy of Sciences, Beijing 100190, China}
\affiliation{Collaborative Innovation Center of Quantum Matter, Beijing, China}

\date{\today}

\begin{abstract}
We performed a comprehensive angle-resolved photoemission spectroscopy study of the electronic band structure of LaOFeAs single crystals. We found that samples cleaved at low temperature show an unstable and highly complicated band structure, whereas samples cleaved at high temperature exhibit a stable and clearer electronic structure. Using \emph{in-situ} surface doping with K and supported by first-principles calculations, we identify both surface and bulk bands. Our assignments are confirmed by the difference in the temperature dependence of the bulk and surface states.
\end{abstract}

\pacs{74.70.Xa, 74.25.Jb, 79.60.-i}

\maketitle

\section{Introduction}

\begin{figure}[!htp]
\begin{center}
\includegraphics[width=0.5\textwidth]{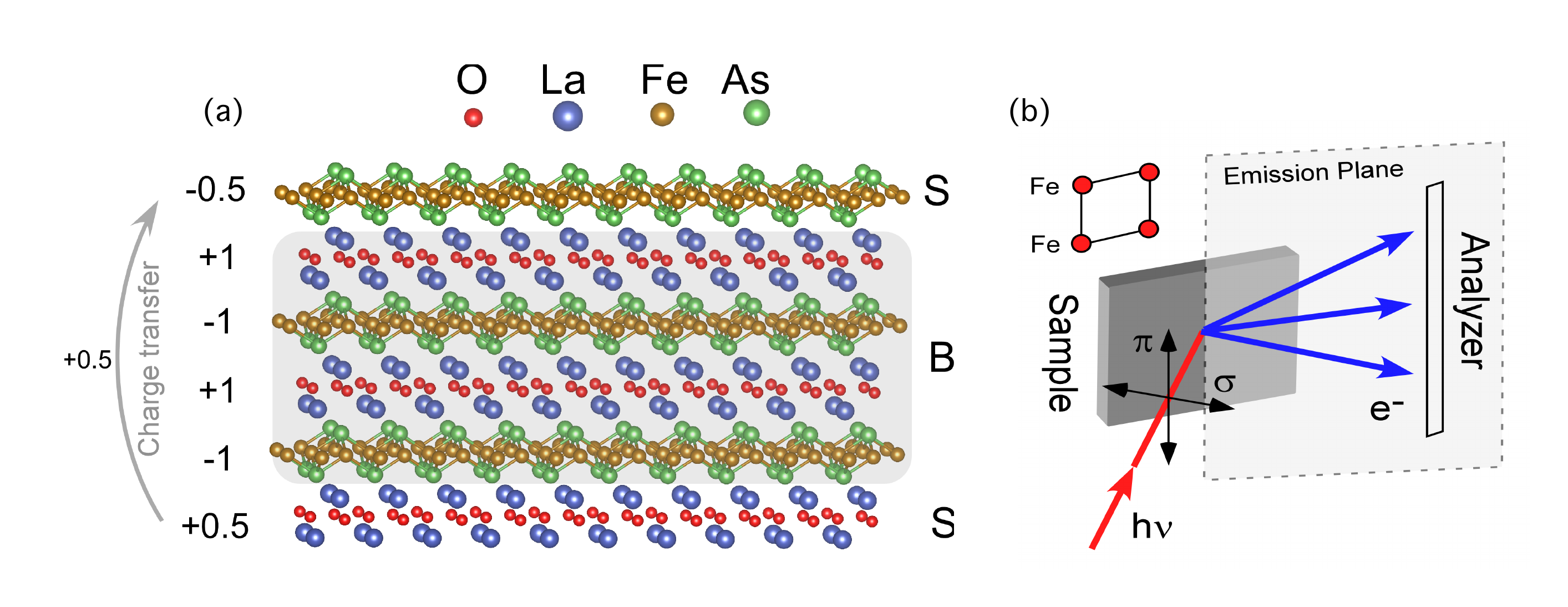}
\end{center}
 \caption{\label{intro} (a) Schematic drawing showing the possible charge transfer on the polarized surface. S and B on the right side stand for surface and bulk components, respectively. (b) ARPES experimental geometry. All the data in this paper are recorded along the $\Gamma$-M (Fe-Fe) direction. The geometries with linear horizontal and linear vertical polarized light are labeled $\sigma$ and $\pi$, respectively.}
\end{figure}

Despite an earlier study\cite{Y_Maeno_Nature372} reporting superconductivity at 5 K in LaOFeP with the so-called 1111 crystal structure, the discovery of a superconducting critical temperature $T_c$ of 26 K in F-doped LaOFeAs\cite{HosonoJACS2008,HosonoNature2008} with the same structure is generally used to mark the beginning of the era of the Fe-based superconductors, and up to now the 1111 family is still the one exhibiting the highest $T_c$'s in bulk single crystals at ambient pressure among all Fe-based superconductors\cite{ChenNature2008,ZhaoCPL2008}. Understanding why high-$T_c$ superconductivity is favored in this system is important but requires a good characterization of its electronic structure. However, the 1111 samples show a polarized cleaved surface that results from both [LaO]$^{+1}$ and [FeAs]$^{-1}$ surface termination layers. This leads to a surface reconstruction after cleaving\cite{NogueraSS1997,NogueraPRB1999}. One possible reconstruction involves a charge transfer from the bottom surface layer to the top surface layer, with a 0.5$e$ charge transfer\cite{NogueraSS1997}, as illustrated in Fig. \ref{intro}(a), which resembles the polar surface issue encountered in YBa$_2$Cu$_3$O$_{7-\delta}$ samples\cite{DamascelliNP2008}. Consequently, the coexistance of surface and bulk electronic states complicates the measurement of the intrinsic electronic structure by angle-resolved photoemission spectroscopy (ARPES)\cite{ShenNature2008,KaminskiPRL2008,LuPC2009,KoepernikPRB2010,FengPRB2010,KaminskiPRB2010,BorisenkoSR2015}. Typical ARPES results always show a large hole pocket around the Brillouin zone (BZ) center ($\Gamma$), although electron pockets around $\Gamma$ have also been reported\cite{FengPRB2010}.


In this paper, we present a detailed investigation by ARPES of the electronic band structure of the parent compound LaOFeAs. We show that the band structure depends on the sample cleaving temperature, possibly due to the surface reconstruction. In particular, high-temperature cleaving produces a stable band structure. Supported by local density approximation (LDA) calculations, we distinguish both surface and bulk states. Upon doping the surface \emph{in-situ} with K, we clearly observed different energy shifts of surface and bulk states on both the core levels and the valence states. We also use temperature-dependent measurements to show that the surface and bulk states evolve differently, thus confirming their assignment. Our results provide a good starting point for extracting the key ingredients responsible for the high $T_c$ of the 1111 ferropnictide superconductors.


\section{Experiment}

High-quality single-crystals of LaOFeAs were grown with NaAs flux. This material shows a structural transition at $T_S=155$ K and a magnetic transition at about $T_S=137$ K \cite{HosonoJACS2008,DaiNature2008}. ARPES measurements were performed at the Advanced Light Source, beamlines BL12 and BL4, using VG-Scienta electron analyzers. The energy resolution was set to 15 meV and the angular resolution was set to 0.2$^\circ$. The experimental geometry is shown in Fig. \ref{intro}(b). All measurements of the valence bands presented in this paper were performed at 80 eV, except mentioned otherwise. Clean surfaces for the ARPES measurements were obtained by cleaving the samples \emph{in situ} in a working vacuum better than $7\times 10^{-11}$ Torr. We use the tight binding model in Ref. [\onlinecite{ScalapinoNJP2009}] with a manual shift of onsite energies and hopping parameters for the $d_{xy}$ and $d_{z^2}$ orbitals to match the experimental data.

\section{Impact of the cleaving temperature}

\begin{figure}[!tb]
\begin{center}
\includegraphics[width=0.5\textwidth]{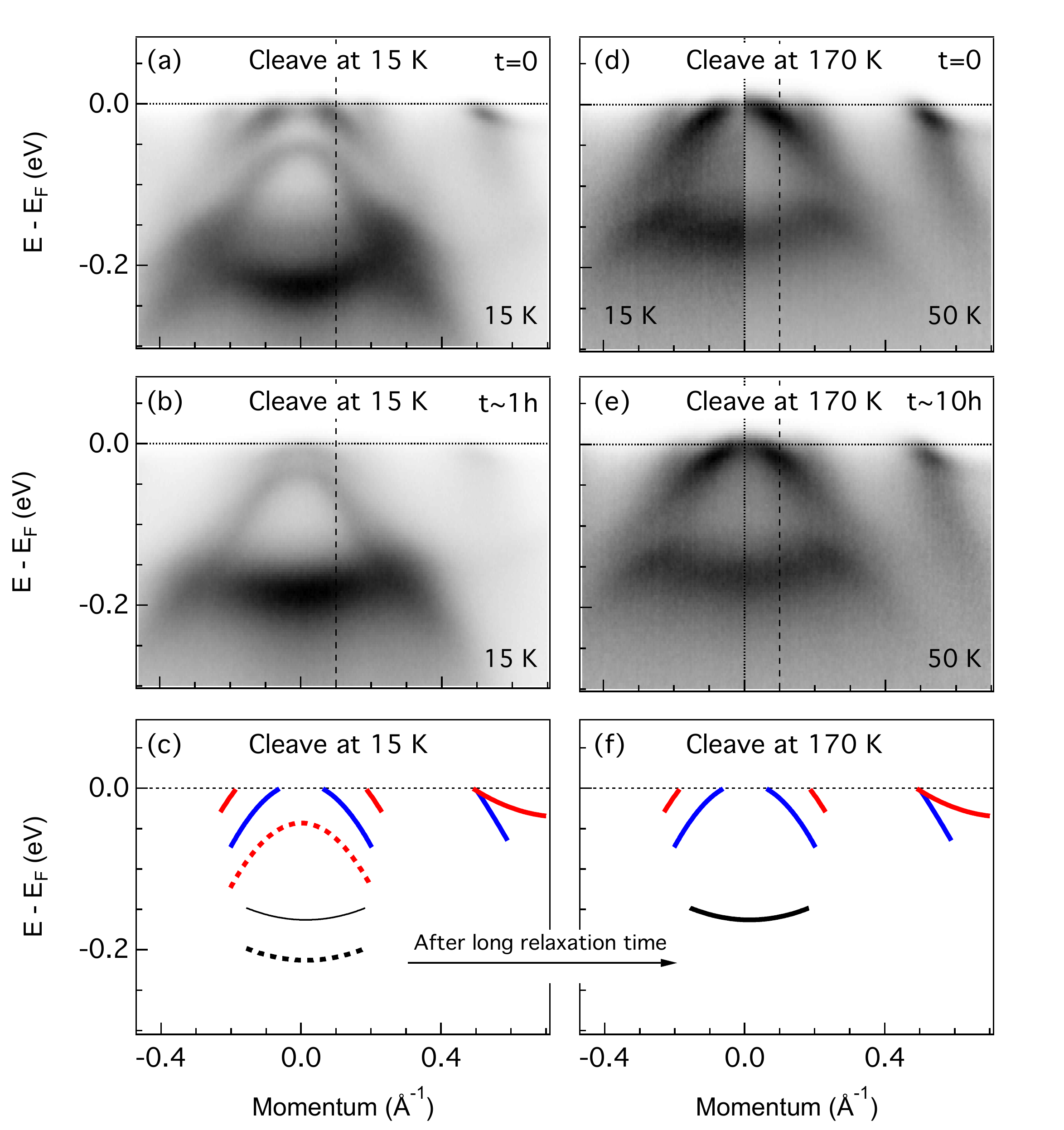}
\end{center}
 \caption{\label{cleavingT} (a) ARPES intensity plot of the band structure of one sample cleaved at 15~K. The data are recorded at 15~K, at the $\Gamma$ point and with $\sigma$ polarization. (b) Same as (a), but recorded one hour later. (c) Cartoon of the band structure of samples cleaved at 15~K. The dashed lines indicate bands very sensitive to time. (d) Same as (a), but with one sample cleaved at 170~K. The data on the left side are recorded at 15~K and the ones on the right side are obtained at 50~K. (e) Same as (d), but recorded 10 hours later, at 50~K. (f) Cartoon of the band structure of samples cleaved at 170~K.}
\end{figure}

We found that the observed band structure of LaOFeAs is strongly affected by the sample cleaving temperature. In Fig. \ref{cleavingT}, we compare the band structure measured on samples cleaved at 15 K and 170 K, and show their evolution with recording time. We notice extra bands for the samples cleaved at 15 K as compared to the samples cleaved at 170 K, as well as band shifts with the recording time. Furthermore, the bands very close to $\Gamma$ in the low-temperature cleaved samples are observed at higher binding energies, which implies an electron doping. However, the outer hole band centered at $\Gamma$ is very large, suggesting over-hole doping. This suggests complicated sets of bands belonging to layers near the surface for the low-temperature cleaved samples. As the surface reconstructs with time, the charge on the polar surface is balanced gradually, and the band structure for the low-temperature cleaved samples tends to be more bulk-like. In contrast, we did not observe any significant change after 10 hours of measurements for the samples cleaved at 170 K. However, we notice a slight band shift between data recorded at 15 K and 50 K, which will be discussed below in Section \ref{t-section}.

The different band structures and their evolution with recording time are sketched in Figs. \ref{cleavingT}(c) and \ref{cleavingT}(f). The band structure of the samples cleaved at 15 K evolves with time towards the one obtained on the samples cleaved at 170 K, indicating that there is a slow reconstruction on the cleaved surfaces that relaxes much faster at high temperature. We conclude that the surfaces obtained by cleaving the samples at high temperature are more representative of the intrinsic properties of LaOFeAs, and hereafter we focus mainly on the samples cleaved at 170 K.


\begin{figure*}[!htb]
\begin{center}
\includegraphics[width=\textwidth]{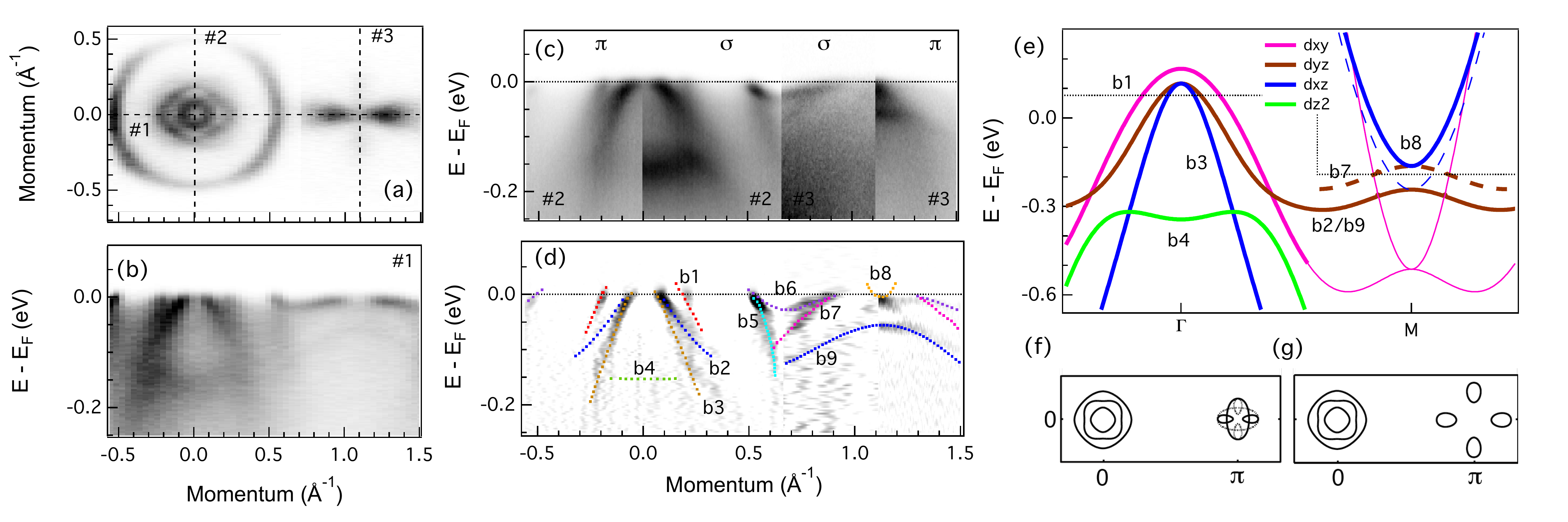}
\end{center}
 \caption{\label{band} (a) FS mapping at 15 K with data from $\sigma$ + $\pi$ polarization. (b) ARPES intensity plot of the band structure along cut \#1 from panel (a). The intensity between $k_x=(0.71, 1.5)$ in (a) is multiplied by 2 to show the details near M. (c) ARPES intensity plots of the band structure along cuts \#2 and \#3, with different polarizations. (d) Curvature intensity plot \cite{curvature} of (c) and sketch of the band structure. (e) LDA calculations, considering the splitting caused by $d$-wave orbital order and contributions from resulting twin domains. The thick lines are the ones observed in the experimental data, whereas the thin lines correspond to bands not observed. The dashed lines refer to bands from a different domain. The horizontal dashed lines correspond approximately to the experimental Fermi level\cite{BrouetPRL2013,BorisenkoNP2016}. (f) - (g) Sketched FSs from LDA calculations with and without orbital ordering and domains, respectively.}
\end{figure*}

\begin{figure*}[!htb]
\begin{center}
\includegraphics[width=\textwidth]{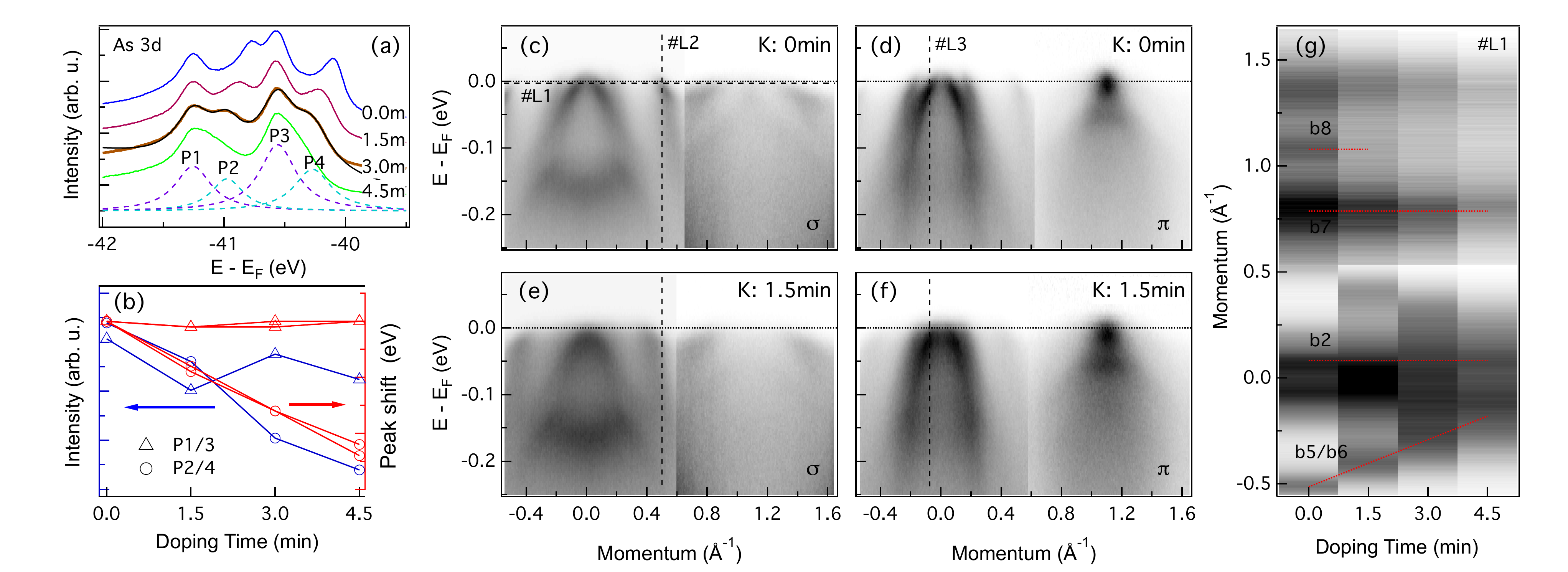}
\end{center}
 \caption{\label{insitu} (a) Core levels of As $3d$ as a function of K doping time. The black line is a fitting of the core levels after 3.0 minutes K doping. The fitting uses a sum of four Lorentzian functions. The dashed purple and light blue lines are the fittings of individual peaks. Due to the spin-orbit coupling, the As $3d$ core levels split into 3$d_{3/2}$ and 3$d_{5/2}$ peaks, with a 5:7 intensity ratio. Thus, we fixed the area ratios P1/P3 and P2/P4 to 5:7. (b) Peak area (blue, left axis) and peak shifts (red, right axis) of P1/P3 and P2/P4, with K doping time. (c) and (d) Band structure recorded before K evaporation at 50 K with $\sigma$ and $\pi$ polarizations, respectively. (e) -(f) Same as (c) - (d), but recorded after K doping. (g) $k_F$ shifts with K doping, at $E = E_F$ (cut \#L1 in (c)). }
\end{figure*}

The full band structure of samples cleaved at 170 K is shown in Fig. \ref{band}. From the FS mapping in Fig. \ref{band}(a), we distinguish one large FS and two small FSs at the $\Gamma$ point, as well as four symmetric hot spots at M. By comparing the intensity plots at different binding energies, we further conclude that the four hot spots are hole-like. From the band structure along high-symmetry lines corresponding to cut \#1, cut \#2 and cut \#3 in Figs. \ref{band}(b) - \ref{band}(d), we distinguish nine bands, which are sketched in Fig. \ref{band}(d). The very large hole FS actually consists of two bands, labeled b5 and b6. These two degenerated hole FSs are too large to account for the bulk doping and should thus be related to surface states induced by the charge transfer on the polar surface, as discussed below. The bands b1, b2 and b3 at $\Gamma$ are consistent with the bulk calculations of the three $t_{2g}$ orbitals, as shown in Fig. \ref{band}(e).

There are two possible ways to explain the four hole-like hot spots at M: i) If we simply ignore nematicity and use the calculations for the tetragonal phase, we obtain the FS shown in Fig. \ref{band}(g)\cite{Jiang_PRB2016}. ii) By taking into account the orbital order\cite{ShenPNAS2011, ZhangPRB2015} and a twin domain structure, we get the FS shown in Fig. \ref{band}(f). Here we used a $d$-wave orbital order\cite{ZhangPRB2015, Jiang_PRB2016} since no splitting is observed at $\Gamma$. The case shown in Fig. \ref{band}(g) can be excluded though since all the FSs are hole-like, which is incompatible with the zero nominal carrier doping of this parent compound. In the second option there is one big electron band at M in a single domain, which can help balancing the carrier concentration. This is the interpretation that we adopt hereafter. However, this electron band at M consists of $d_{xy}$ orbital character, which has bad coherence at M, and thus it is not present in the experimental data, as in the case of FeSe single crystals\cite{ColdeaPRB2015,ZhangPRB2015}. Bands b9 and b7 should be $d_{yz}$ bands from different domains, and they are part of b2. Band b8 should carry a $d_{xz}$ orbital character from one domain. As with the $d_{xy}$ band, the other $d_{xz}$ band from the different domain at M is not observed.

We mark the observed bands in the LDA calculations in Fig. \ref{band}(e). The bands not observed by experiments are drawn with thin lines. If we count the FS volume of the two degenerate hole FSs (b5/b6), we find $2\times 0.35$e/Fe using the Luttinger theorem, which is much larger than the maximum charge transfer 0.5e/Fe induced by the polar surface, as shown in Fig. \ref{intro}(a). We speculate that the two hole FSs may come from different layers, or that there are some unobserved surface electron bands\cite{FengPRB2010}.

\section{\emph{In-situ} K doping}

Evaporating K \emph{in-situ} on the surface of samples is a technique that has been proved useful to electron-dope surfaces of high-temperature superconductors\cite{DamascelliNP2008,ZhangAPL105}. It has also been used to kill some surface states in the Fe-based superconductors\cite{RichardJPCM2014}. We thus apply this technique on LaOFeAs in order to help us distinguishing the surface and bulk components of the measured electronic structure. We first investigate the core level shifts with K doping time. Since the chemical environments at the surface and in the bulk are different due to the polar surface of LaOFeAs, the surface and bulk core levels should be different. Indeed, we see in Fig. \ref{insitu}(a) that the As 3$d_{3/2}$ and As 3$d_{5/2}$ core levels split into four peaks, labeled from P1 to P4. For a quantitative understanding, we fit the peaks with Lorentzian functions. The extracted peak positions and areas are shown in Fig. \ref{insitu}(b). We find that P1 and P3 on one hand, and P2 and P4 on the other hand, form two pairs with a fixed peak intensity ratio, suggesting that one set of peaks belongs to the surface states whereas the other pair is associated with the bulk. While the positions of the P1 and P3 peaks do not change with K doping time, the positions of the P2 and P4 peaks move to higher binding energy, in agreement with an electron doping. Since surface states are more sensitive to \emph{in-situ} K doping, we assign the P2 and P4 peaks to surface states, and attribute the P1 and P3 peaks to bulk states.

Next we use K doping to study the evolution of the valence states. The band structures before and after K doping are shown in Figs. \ref{insitu}(c) - \ref{insitu}(d) and Figs. \ref{insitu}(e) - \ref{insitu}(f), respectively. With the help of lines \#L2 and \#L3 (see Figs. \ref{insitu}(c) and \ref{insitu}(d) for the momentum locations), we found that except for the broadening caused by the disordered K atoms, the most obvious change is the large downward shift of the b5/b6 bands, while other bands have no obvious shifts. To show the band change clearly, we display the MDC change at line \#L1 (see Fig. \ref{insitu}(c) for the energy location) with K doping time in Fig. \ref{insitu}(g). A common feature in this panel is the weakening of intensity with K doping due to the disorder introduced by the K atoms. Nevertheless, our results show clearly that while the $k_F$ position of the b5/b6 bands shifts significantly with surface doping, the $k_F$ positions of the b2, b7 and b8 bands almost do not. As we did with the core levels, this observation indicates that the b5/b6 bands are from the surface, while the others are more bulk-representative.

\section{Temperature dependence}
\label{t-section}

\begin{figure}[!t]
\begin{center}
\includegraphics[width=0.45\textwidth]{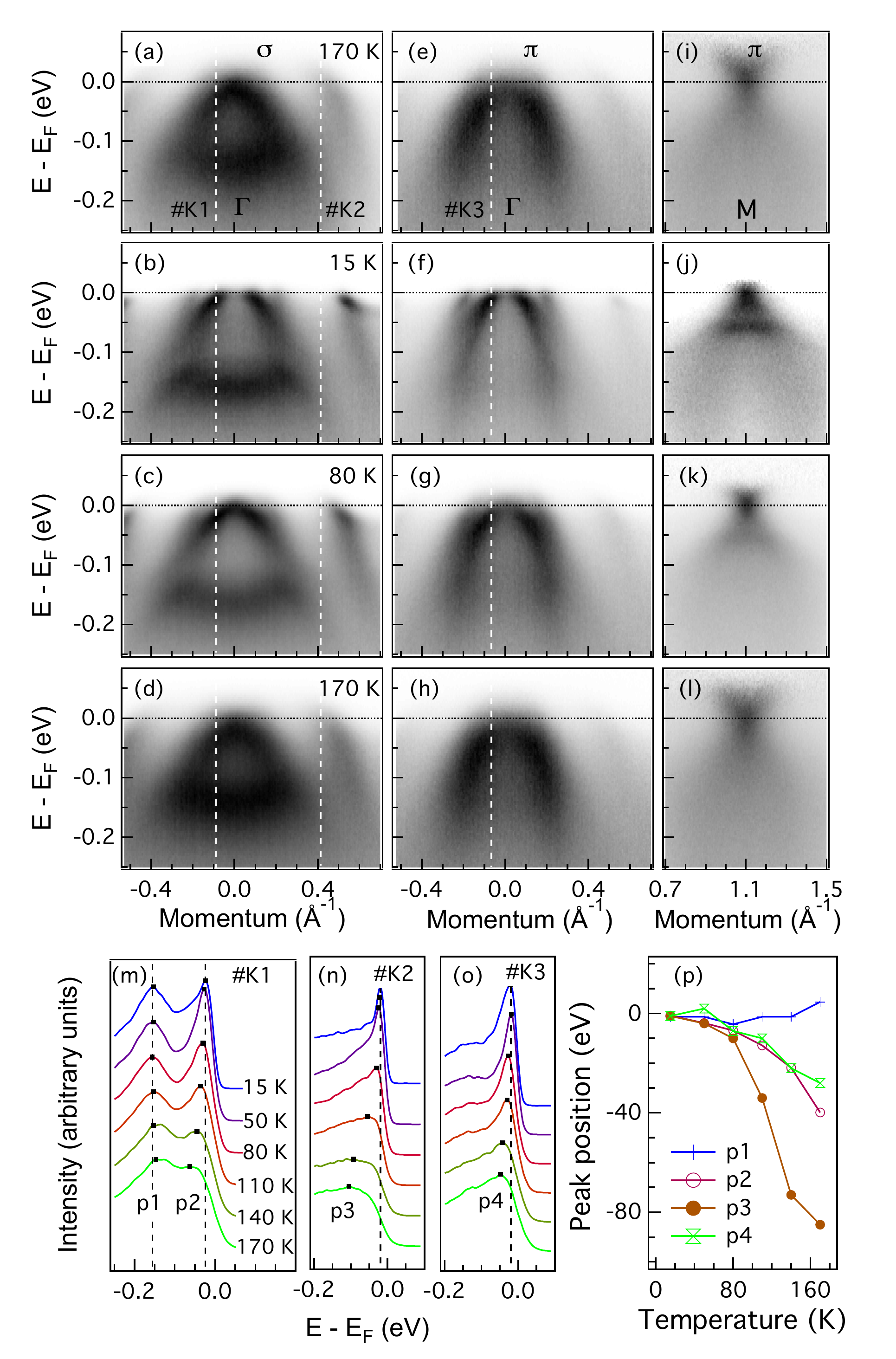}
\end{center}
 \caption{\label{temperature} (a) - (d) ARPES intensity plots recorded at $\Gamma$ with $\sigma$ polarization, for the temperature cycle 170~K$\rightarrow$15~K$\rightarrow$80~K$\rightarrow$170~K. (e) - (h) same as (a) - (d), but with $\pi$ polarization. (i) - (l) Same as (e) - (h), but for a cut at M and divided by the Fermi Function at the corresponding temperature. (m) - (o) Temperature evolution of the EDCs at momenta \#K1, \#K2 and \#K3, respectively. Black markers indicate the peak positions. (p) Peak position shifts as a function of temperature. Labels corresponding to the peaks showed in panel (m) - (o).}
\end{figure}

From the temperature evolution of the band structures displayed in Fig. \ref{temperature}, we notice obvious band shifts, as reported previously for other Fe-based superconductors\cite{BrouetPRL2013,KaminskiPRL2013,PH_LinPRL111}. The band shifts cannot be attributed to the aging effect since we checked the band structure at 50 K (Fig.\ref{cleavingT}) and 170 K (Fig.\ref{temperature}) after a thermal cycle and found no shift. Interestingly, the band shifts of the surface and bulk states with temperature are different. The outer hole bands b5/b6 at $\Gamma$ in Figs. \ref{temperature}(a) - \ref{temperature}(d), which we attribute to surface states, show a large shift towards the high binding energies, whereas the bulk hole band b2 in Figs. \ref{temperature}(a) - \ref{temperature}(d) and the hole band b3 in Figs. \ref{temperature}(e) - \ref{temperature}(h) show a relatively small shift.


We display the EDCs at the positions labeled \#K1, \#K2 and \#K3 at different temperatures in Figs. \ref{temperature}(m) - \ref{temperature}(o). The bulk bands b2 and b3 have a shift of 50$\sim$60 meV, while the surface band b5 has a shift of about 110 meV, about twice as large. The band shifts can be partially explained by the carrier conservation and the large decrease in the density-of-states decrease near the Fermi level (hole band top) \cite{BrouetPRL2013}. In Figs. \ref{temperature}(i) - \ref{temperature}(l) we show the band structure at different temperatures at the M point. Under $\pi$ polarization, the most obvious feature is the electron band b8, which is very strong and thus has a large intensity even below E$_F$. The band structure under $\sigma$ polarization is too weak [Figs. \ref{insitu}(c) and \ref{insitu}(e)] for the temperature dependent study and thus ignored.

\section{Summary}

In summary, we performed a detailed study of the band structure of the parent compound LaOFeAs. We identified both surface and bulk-representative states by cleaving the samples at 170 K. By doping the surface of LaOFeAs \emph{in-situ} using K, we confirmed our classification of the surface and bulk states for the valence bands, and further proved this result by looking at their temperature evolution. Our results indicate a routine method for disentangling bulk and surface states in the Fe-based superconductors.

We acknowledge D. Chen, X. Shi, S.F. Wu for useful discussions. This work was supported by grants from CAS (XDB07000000), MOST (2015CB921300, 2011CBA001000, 2013CB921700, 2012CB821400), NSFC (11474340, 11274362, 11234014, 11190020, 91221303, 11334012, 11274367, 11474330). The Advanced Light Source is supported by the Director, Office of Science, Office of Basic Energy Sciences, of the U.S. Department of Energy under Contract No. DE-AC02-05CH11231.

\bibliographystyle{apsrev}
\bibliography{biblio_LaOFeAs}

\end{document}